\documentclass[10pt, twocolumn]{article}
\usepackage[utf8]{inputenc}
\usepackage[english]{babel}
\usepackage{amssymb}
\usepackage{amsmath}

\usepackage{lineno}

\usepackage[margin=0.75in, top=0.75in]{geometry}

\usepackage{algorithm}
\usepackage[noend]{algpseudocode}

\usepackage{float}

\floatstyle{boxed}
\newfloat{deffloat}{h}{deffloats}
\floatname{deffloat}{Definitions}

\usepackage{empheq}

\usepackage{framed} 
\usepackage[framed]{ntheorem}
\theoremstyle{nonumberplain}
\newframedtheorem{frm-thm}{}

\usepackage{multicol}
\usepackage{sectsty}
\sectionfont{\bfseries\Large\raggedright}
\sectionfont{\bfseries\large\raggedright}

\usepackage{graphicx}
\graphicspath{ {./images/} }
\usepackage[export]{adjustbox}
\usepackage{subcaption}

\usepackage[
backend=biber,
style=alphabetic,
citestyle=authoryear
]{biblatex}
\usepackage[colorlinks,citecolor=blue,urlcolor=blue,bookmarks=false,hypertexnames=true]{hyperref}

\author{
  Jeffrey Wong\thanks{Wong was employed at Netflix when this work began.} \\
  Computational Causal Inference
}

\title{Delta Vectors Unify the Computation for Linear Model Treatment Effects}
\date{October 12, 2022}

\begin{document}

\maketitle

\section{Introduction}

The science of cause and effect is extremely sophisticated and extremely hard to scale.
Across experimentation causal inference is used to evaluate treatments, and execute decisions. Different fields of data science, such as economics, psychology, marketing science, statistics, and computer science, are contributing to information that can be extracted out of randomized controlled experiments. These scientists get rich insights by analyzing global effects, effects in different segments, and trends in effects over time. They use propensity scores to project external validity. To support the analysis of relative effects, scientists derive challenging ratio distributions. These mathematical models support the evaluation of an experiment. Afterwards, a decision to roll out needs to be made. Even in this space, scientists are inventing new methods to determine an optimal roll out policy; sometimes the decision does not concern the magnitude of the effect, it simply asks for a probabilistic ranking of the options. All of these modes of analysis are mathematically advanced, but are hard to implement in a software platform.

While the analytical capabilities in experimentation are advancing, we require new innovation within engineering and computational causal inference to enable an experimentation platform to make these analyses performant and scalable (\cite{wong2020computational}). Of significant importance: we must unify the computing strategy for these models so that they can be consistently applied across experiments. In doing so, the industry can make significant progress towards developing a flywheel that unifies and accelerates the evaluation and roll out of experiments. In order to support unified computation, this paper introduces baseline vectors and delta vectors as common structure for linear model treatment effects. Assuming a linear model can be fit, this structure allows us to incorporate arbitrary covariates into most treatment effect problems without increasing engineering complexity. It also creates a simple way to derive the standard error for arbitrarily complex heterogeneous treatment effects and time varying effects. Finally, we are also able to derive relative effects, and rank treatment effects with a Bayesian linear model. Baseline vectors and delta vectors are widely applicable to linear models, including the glms, linear models with regularization, and Bayesian linear models, and are a crucial underpinning for the engineering of linear model treatment effects.

\section{Background}

First, we assume a small linear model with the form

$$y = \alpha + X\beta_1 + W\beta_2 + (X \cdot W) \beta_3 + \varepsilon$$
where $X$ is a matrix of covariates, $W$ is a vector for the treatment indicator, and $(X \cdot W)$ is the interaction.
\cite{gelman} uses this model to discuss an educational experiment where they study the effect of an educational television program on reading scores in elementary school students. The analysis makes use of covariates, such as pre-test scores and grade, to decrease variance on the treatment effect. It was also hypothesized that the treatment effect itself is heterogeneous with pre-test scores.

Standard regression software fits the linear model and outputs a summary of the regression report. However, such software requires manually computing heterogeneous effects, and is not scalable for an experimentation platform that has to process arbitrary covariates. We demand a unified compute strategy that can estimate heterogeneous effects along multiple covariates that is also reuseable for average effects, time varying effects, relative effects, and ranking of effects.
In the report below, the linear model is fit with an interaction to measure heterogeneous effects along pre test scores.

\includegraphics[width=\columnwidth]{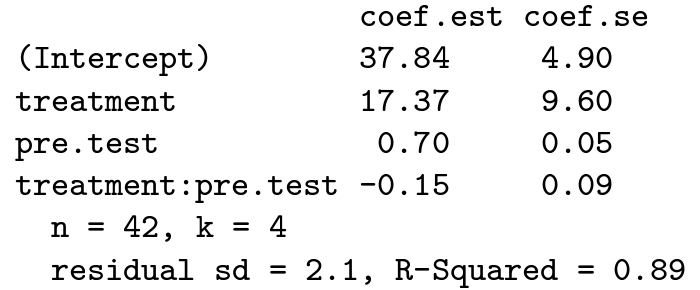}

We may think it is also important to measure heterogeneity along other covariates, such as grade, family background, number of siblings, and the school. It is not easy to read multiple heterogeneous effects that are derived from multiple interaction terms. First, the linear algebra needs to be derived for each heterogeneous effect using select main effects and interaction effects. Second, the relevant covariances need to be identified, pulled, multiplied, and added to derive the standard error. This is difficult to codify in software, especially since not all main effects or interactions are relevant to the treatment effect.

Writing manual functions to combine coefficients and covariances, as is described in \cite{gelman}, is not a scalable platform solution. Below, our solution using baseline vectors and delta vectors not only makes it easy to derive and execute the linear algebra for heterogeneous effects with arbitrary amounts of interaction terms, it is also a reuseable strategy for time dynamic effects, relative effects, and ranking of effects. Furthermore, it is also compatible with computationally efficient methods for fitting the model, such as conditionally sufficient statistics (\cite{wong2021you}). This creates a completely optimized computational stack.

\section{Baseline Vectors and Delta Vectors}

Baseline and delta vectors are two primitives that underpin the unified compute strategy; these primitives can be implemented in any software environment with just multiplication and addition. They also make derivations of statistical properties simple, opening opportunities in causal inference to be interesting and scalable. The \textbf{baseline vector} for treatment arm $W = w$ is a specifically constructed row vector, $B(W = w)^\top$, such that multiplication with parameters $\hat{\beta}$ produces
the average value of $y$ under treatment arm $w$.

$$B(W = w)^\top \hat{\beta} = \frac{1}{n} \sum_i y^{(w)}_i.$$

When considering an experiment with a treatment arm $W = 1$ and a control arm $W = 0$,
the average treatment effect can be rewritten as the difference between two baseline vectors.
This difference in baseline vectors motivates the construction of the \textbf{delta vector}, $D(W_2 = 1, W_1 = 0)^\top$, a similar row vector such that

\begin{align*}
    ATE(W_2 = 1, W_1 = 0) &= \frac{1}{n} \sum_i y^{(1)}_i - y^{(0)}_i \\
        &= B(W = 1)^\top \hat{\beta} - B(W = 0)^\top \hat{\beta} \\
        &= D(W_2 = 1, W_1 = 0)^\top \hat{\beta}.
\end{align*}

These two vectors can be implemented using just multiplication and addition
in any software. The baseline vector computes the column means of $X$, then constructs
$
\begin{bmatrix}
1 & \bar{X} & w & (\bar{X} \cdot w)
\end{bmatrix}.$ The delta vector, $D(w_2, w_1)^\top$, can be computed as the difference in
baseline vectors, or it can be directly computed as 
$
\begin{bmatrix}
0 & 0 & w_2 - w_1 & (\bar{X} \cdot (w_2 - w_1))
\end{bmatrix}.$

In addition, the statistics for causal inference have
easily derived properties. In the case of the baseline,
or average of a single potential outcome,
the variance is simply a multiplier on the variance of
$\hat{\beta}$. Likewise, the variance on the average treatment effect is simply $D^T \text{cov}(\hat{\beta}) D$.

\section{Absolute Effects}

Other than average treatment effects, we can also compute
conditional average treatment effects (CATE), and time dynamic treatment effects (DTE) easily. We use delta vectors to derive the expectation of the effect, as well as its variance, even with an arbitrary list of covariates. This derivation is simple, and does not require manually picking coefficients or covariances that should be combined, yielding a strategy that is simple for users and is easy to codify.

The conditional effect, $CATE(X = x, W_2 = w_2, W_1 = w_1)$, is the average treatment effect computed over
the subset of observations with feature $X = x$. Again, using multiplication and addition, we compute the delta vector 
$$D(X = x, w_2, w_1)^\top = \begin{bmatrix}
0 & 0 & w_2 - w_1 & (x \cdot (w_2 - w_1))
\end{bmatrix}$$
and multiply by $\hat{\beta}$ to get $CATE(X = x)$. This estimator is a simple multiplication with a random variable, so its variance is easily derived as $D(x)^T \text{Cov}(\hat{\beta}) D(x).$ Although we compute a conditional effect, we do not
need to identify specific coefficients or covariances to combine, we simply operate on the entirety of $\hat{\beta}$ and $\text{Cov}(\hat{\beta})$.

In many applications we also want to measure heterogeneity in the effect, or how the effect for a group $X = x$ is different than all other groups $X = -x$. The heterogeneous effect is

\begin{align*}
HTE(x, w_2, w_1) &= CATE(x, w_2, w_1) - CATE(-x, w_2, w_1) \\
  &= [D(x, w_2, w_1) - D(-x, w_2, w_1)]^\top \hat{\beta}
\end{align*}
Again, through the elegance of delta vectors this can be computed efficiently. The variance of this estimator is $[D(x, w_2, w_1) - D(-x, w_2, w_1)]^\top \text{Cov}(\hat{\beta}) [D(x, w_2, w_1) - D(-x, w_2, w_1)].$

Time varying effects are a specific type of heterogeneity and can be computed in the same way above, with the exception that $\text{Cov}(\hat{\beta})$ should account for within observation correlation.

\section{Relative Effects}

Average, conditional, and heterogeneous effects measure absolute differences.
We may be interested in the average relative difference,

$$ARE = \frac{\sum_i y^{(w_2)}_i}{\sum_i y^{(w_1)}_i} - 1.$$
\cite{deng2018applying} applied the delta method to derive a compute strategy and variance estimate for the ARE when there are no covariates. Baseline and delta vectors can expand the strategy when the ARE is derived from a linear model with covariates.
 
The expectation and variance of a ratio estimator $\frac{R}{S}$ under a second order Taylor expansion is found in \cite{van2000mean}.

\begin{align*}
    E(R/S) &= \frac{E(R)}{E(S)} - \frac{\text{Cov}(R, S)}{E(S)^2} + \frac{Var(S)E(R)}{E(S)^3} \\
    Var(R/S) &= \frac{E(R)^2}{E(S)^2} 
    [\frac{\text{Var}(R)}{E(R)^2} - 2\frac{Cov(R, S)}{E(R)E(S)} + \frac{\text{Var}(S)}{E(S)^2}]
\end{align*}

In the simple case where $R$ is the baseline estimate for the treatment group, and $S$ is the baseline estimate for the control group, neither of which are derived with covariates then $\text{Cov}(R,S) = 0$ due to independence. This reduces the moments to those found in \cite{deng2018applying}. In a linear model with covariates $X$ that exist for both the treatment and control groups, $\text{Cov}(R,S) \neq 0$. Nonetheless, using baseline and delta vectors we can easily derive the sampling distribution for the ARE as

\begin{align*}
    ARE &= \frac{\sum_i y^{(w_2)}_i}{\sum_i y^{(w_1)}_i} - 1 \\
        &= \frac{D(w_2, w_1)^\top \hat{\beta}}{B(w_1)^\top \hat{\beta}} \\
    E(R) &= D(w_2, w_1)^\top \hat{\beta} \\
    E(S) &= B(w_1)^\top \hat{\beta} \\
    Var(R) &= D(w_2, w_1)^\top \text{Cov}(\hat{\beta}) D(w_2, w_1) \\
    Var(S) &= B(w_1)^\top \text{Cov}(\hat{\beta}) B(w_1) \\
    Cov(R, S) &= D(w_2, w_1)^\top \text{Cov}(\hat{\beta}) B(w_1).
\end{align*}

\section{Probabilistic Ranking with a Posterior Distribution}

Bayesian linear models also benefit from baseline and delta vectors. Assume
that a Bayesian linear model has been fit so that there is a posterior distribution $p(\beta | \text{data})$, with CDF $\text{P}(\beta | \text{data})$.

In a two arm experiment, the posterior distribution can be used to answer a relevant question: ``What is the probability
that the treatment is better than the control?" Reframing the question to ``What is the probability that the treatment effect is positive?" and extending for a specific
set of observations with $X = x$ yields yet another application of delta vectors. It is equal to measuring

$$\text{Prob}(D(x, w_2, w_1)^\top \beta > 0).$$

Since the delta vector is fixed and $\beta$ is the only random variable, we can reparametrize another distribution, $P'$, for the quantity $D(x, w_2, w_1)^\top \beta$ so that $$\text{Prob}(D(x, w_2, w_1)^\top \beta > 0) = 1 - P'(0).$$

If the posterior of the regression parameters is normally distributed, for example by using a normal prior with a fixed variance as in \cite{williams2006gaussian}, or by using a Laplace approximation (\cite{tierney1986accurate}), then the probability that the treatment is better than the control can be computed efficiently. Let $\Phi(x, \mu, \Sigma)$ be the CDF of a normal distribution with mean $\mu$ and covariance $\Sigma$.
The treatment effect will also be normally distributed with mean $\hat{\mu} = D(x)^\top E[\beta | \text{data}]$ and covariance $\hat{\Sigma} = D(x)^\top \text{Cov}(\beta | \text{data})D(x)$, so the probability is 

$$\text{Prob}(D(x, w_2, w_1)^\top \beta > 0) = 1 - \Phi(0, \hat{\mu}, \hat{\Sigma}).$$

In a three arm experiment, another relevant question is: ``What is the probability that arm $w_2$ is the best arm?" In this case, we are not overly concerned with the magnitude of the effects; we aim to simply pick the best choice among the arms taking into consideration the effect size and its uncertainty. Reframing, we ask ``What is the probability that the treatment effect of $w_2$ on $w_1$ is positive and the treatment effect of $w_2$ on $w_3$ is also positive?" This probability for the joint event can be measured as 

$$\text{Prob}(
\begin{bmatrix}
 D(x, w_2, w_1)^\top \beta \\
 D(x, w_2, w_3)^\top \beta
\end{bmatrix} >
\begin{bmatrix}
 0 \\
 0
\end{bmatrix}
).
$$

Assuming the posterior on $\beta$ is normal, the joint event
$
\begin{bmatrix}
 D(x, w_2, w_1)^\top \beta \\
 D(x, w_2, w_3)^\top \beta
\end{bmatrix} >
\begin{bmatrix}
 0 \\
 0
\end{bmatrix}
$ is multivariate normal with mean 

\begin{align*}
  \hat{\mu} &= \begin{bmatrix}
            D(x, w_2, w_1)^\top E[\beta | \text{data}] \\
            D(x, w_2, w_3)^\top E[\beta | \text{data}]
         \end{bmatrix} \\
  \hat{\Sigma} &= \begin{bmatrix}
             D(x, w_2, w_1)^\top \\ D(x, w_2, w_3)^\top
             \end{bmatrix}
             \text{Cov}(\beta | \text{data})
             \begin{bmatrix}
             D(x, w_2, w_1) & D(x, w_2, w_3)
             \end{bmatrix}
\end{align*}

Finally, the probability that arm $w_2$ is the best among three choices is
$$1 - \Phi(0, \hat{\mu}, \hat{\Sigma})$$

\section{Conclusion}

We have shown that baseline vectors and delta vectors create a unified compute strategy for
average effects, conditional effects, heterogeneous effects, relative effects, and ranking of effects when using a linear model. The compute strategy is widely compatible with any model that has a linear form, including regularized linear models, econometrics models and Bayesian linear models. This greatly simplifies the engineering of an experimentation platform that needs to employ these mathematical models with arbitrary many covariates. This paper assumed that the linear model has already been fit and the parameters $\hat{\beta}$ have been estimated; however baseline vectors and delta vectors are also compatible with efficient strategies for fitting the model such as conditionally sufficient statistics (\cite{wong2021you}). This creates an extremely efficient software stack. In addition to aiding the engineering of large scale causal inference systems, baseline vectors and delta vectors also made the derivation of effects using linear algebra clearer. 

\printbibliography

\end{document}